\documentclass{article}
\usepackage{spconf,amsmath,graphicx,hyperref}
\usepackage[table]{xcolor}
\usepackage{multirow}
\usepackage{afterpage} 
\usepackage{cite}
\usepackage{float}  
\usepackage{booktabs} 
\usepackage[english]{babel}
\usepackage{csquotes}
\usepackage{subcaption}
\usepackage{enumitem}

\title{AFT: An Exemplar-Free Class Incremental Learning Method for Environmental Sound Classification}
\name{Author(s) Name(s)\thanks{Thanks to XYZ agency for funding.}}
\address{Author Affiliation(s)}
\name{Xinyi Chen$^{\star}$ \qquad Xi Chen$^{\dagger}$ \qquad Zhenyu Weng$^{\star}$ \qquad Yang Xiao$^{\ddagger}$}
\address{
$^{\star}$South China University of Technology, China\\
$^{\dagger}$The Chinese University of Hong Kong, Shenzhen, China\\
$^{\ddagger}$The University of Melbourne, Australia
}

\begin{document}
\ninept
\maketitle
\begin{abstract}
As sounds carry rich information, environmental sound classification (ESC) is crucial for numerous applications such as rare wild animals detection. However, our world constantly changes, asking ESC models to adapt to new sounds periodically. The major challenge here is catastrophic forgetting, where models lose the ability to recognize old sounds when learning new ones. Many methods address this using replay-based continual learning. This could be impractical in scenarios such as data privacy concerns. Exemplar-free methods are commonly used but can distort old features, leading to worse performance. To overcome such limitations, we propose an Acoustic Feature Transformation (AFT) technique that aligns the temporal features of old classes to the new space, including a selectively compressed feature space. AFT mitigates the forgetting of old knowledge without retaining past data. We conducted experiments on two datasets, showing consistent improvements over baseline models with accuracy gains of 3.7\% to 3.9\%.
\end{abstract}

\section{Introduction}
\label{sec:intro}

Environmental sound classification (ESC)~\cite{piczak2015esc} is an important task in audio signal processing. Its goal is to help computers recognize and label sounds in daily life, such as animal calls or crowd chatter. Compared with speech-related tasks like speech recognition or speaker identification, environmental sounds do not carry clear meaning or speaker identity cues~\cite{10.1016/j.asoc.2024.112619}. As a result, classification mainly depends on acoustic features. At the same time, the sound environment is not fixed and changes constantly, which makes ESC models face new and unseen classes of sounds~\cite{yin2025exploring,xiao2024mixstyle,xiao2024wilddesed}. The need to adapt to completely new acoustic classes introduces the main challenge of catastrophic forgetting, where models lose the ability to recognize previous sounds once they learn new ones.

To solve the problem of catastrophic forgetting, recent work has focused on class incremental learning (CIL)~\cite{masana2022class}, which is the most practical setting in continual learning \cite{wang2024comprehensive,zhuang2022acil,pandey2024class,xiao25c_interspeech,zhuang2024ds}. In CIL, models learn new classes step by step while retaining old classes' knowledge. Existing CIL methods for sound classification can be roughly divided into two categories based on their approach to handling historical data. The first category is replay-based methods~\cite{xiao2022continual,wang2019continual,xiao2024ucil,yang2023dual,rk,yang2024improving, peng2024dark}, which retain knowledge of old classes by replaying a portion of historical data, but accessing this data poses challenges due to privacy and security considerations. To address this limitation, the second category consists of exemplar-free methods like regularization~\cite{zhang2024remember}, knowledge distillation~\cite{berg2021continual}, analytic learning~\cite{yue2024mmal}, or dynamic network expansion~\cite{xiao2024configurable}, which eliminate the need to store any historical data. ~\cite{10447952} keeps old knowledge by minimizing the differences in feature representations and logits between learners, while ~\cite{zhang2024remember} enables the model to adaptively optimize gradient directions to capture the balance between new and old knowledge. Such methods are better suited to memory-constrained edge devices and mitigate data privacy concerns, as no past audio samples need to be stored. 

\begin{figure}[t!]
  \centering
  \begin{subfigure}[t]{0.49\linewidth}
    \includegraphics[width=\linewidth]{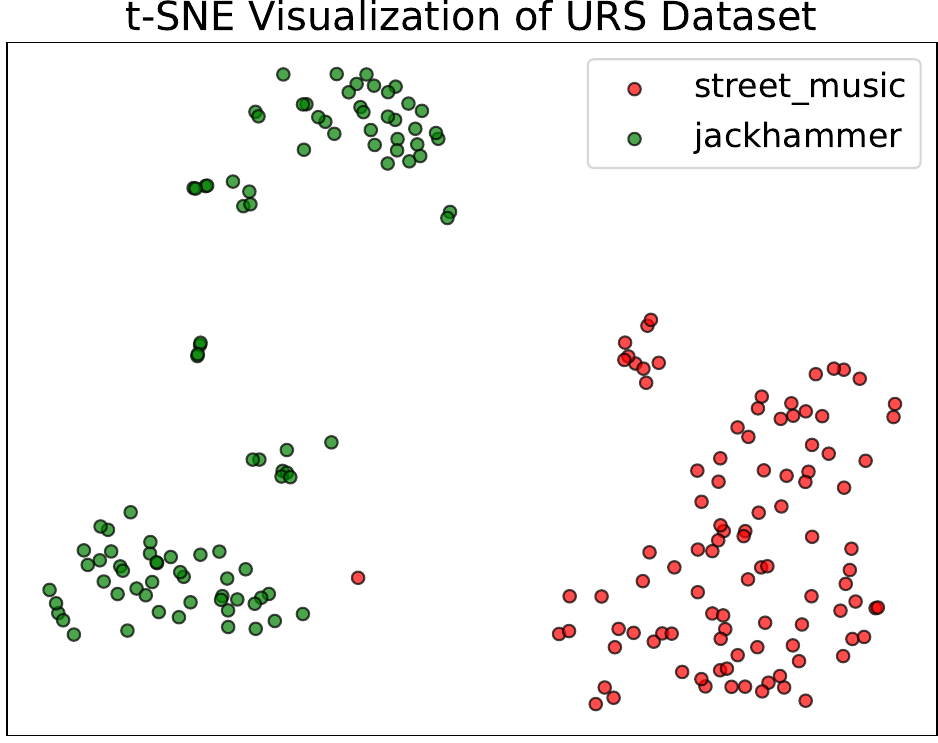}
    \caption{}\label{fig:before}
  \end{subfigure}\hfill
  \begin{subfigure}[t]{0.49\linewidth}
    \includegraphics[width=\linewidth]{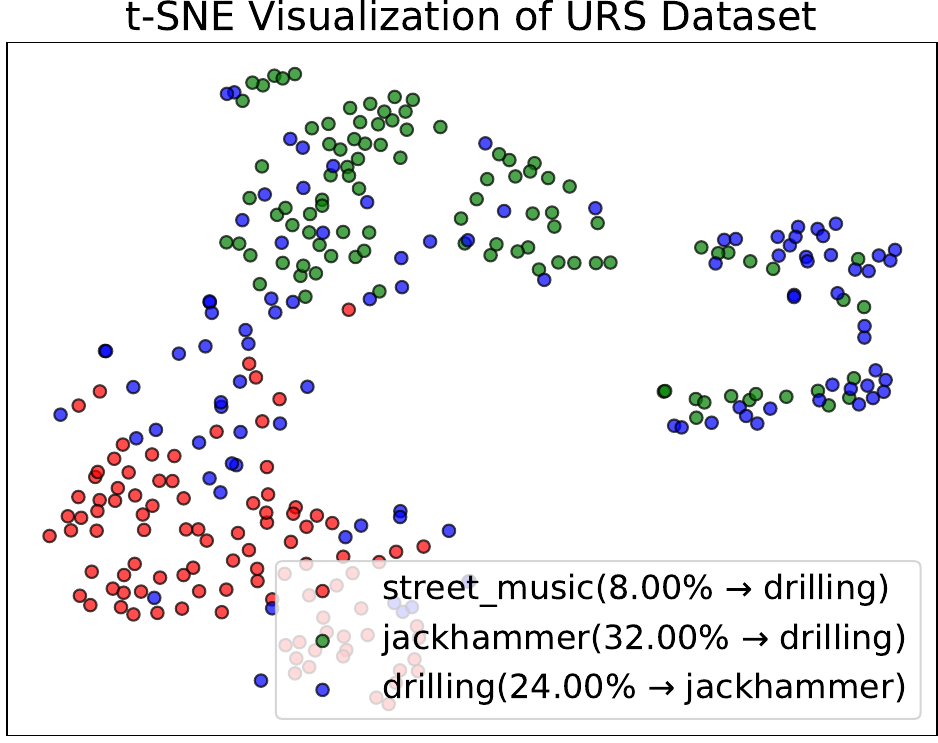}
    \caption{}\label{fig:after}
  \end{subfigure}

  \caption{(a) presents t-SNE feature clusters of street music and jackhammer (trained in the previous phase) with clear separation;
   (b) shows feature distribution after fine-tuning only on drilling data: jackhammer is 32\,\% mis-classified as drilling, whereas street music maintains a relatively clear feature boundary.}
  \label{fig:misalign}
  \vspace{-3mm}
\end{figure}

However, current exemplar-free methods still face performance degradation when novel and existing sound classes share high acoustic similarity (e.g., in frequency patterns or temporal stability). Without historical data as a reference, learning new classes gives rise to a critical issue: the temporal features of old classes become severely misaligned within the updated feature space. This problem is particularly noticeable for acoustically similar sounds. For example, \enquote{drilling} and \enquote{jackhammer} both have short, sharp high-frequency components, whereas \enquote{street music} has a more slow-paced temporal profile. As the model updates its weights to learn new acoustic characteristics, the features of old classes drift away from their previously learned feature space. This eventually leads to the loss of key cues that help recognize sounds the model learned before, as shown in the Figure~\ref{fig:misalign}: \enquote{jackhammer} is 32\% misclassified as \enquote{drilling}, while \enquote{street music}, which has clearly different temporal features, maintains relatively high recognition accuracy.

\begin{figure*}[htbp]
  \centering
  \includegraphics[width=1\textwidth]{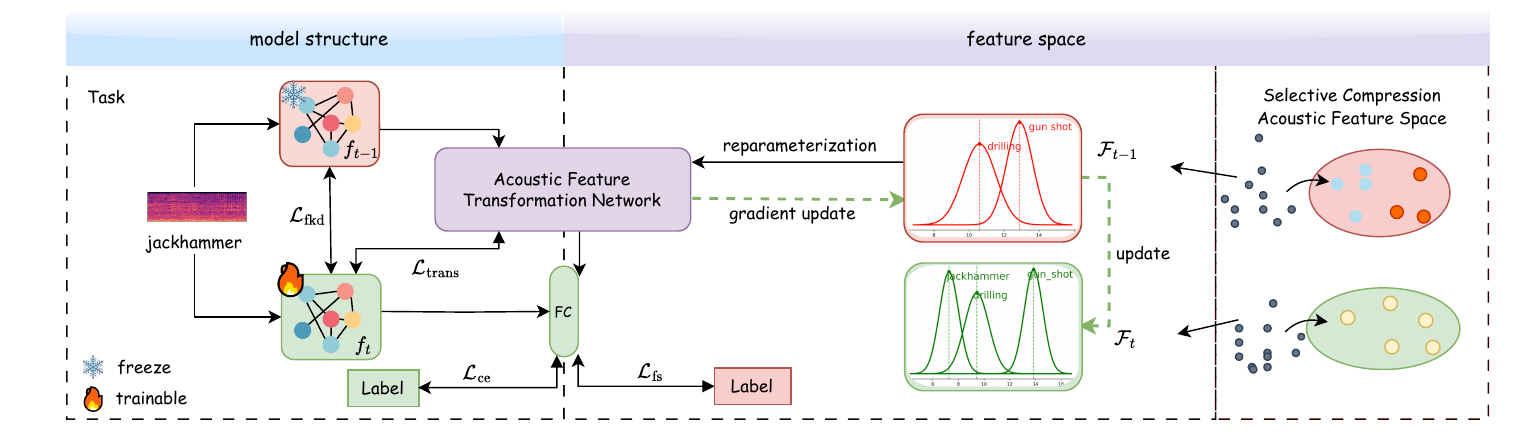}
  \vspace{-2mm}
  \caption{The diagram shows the AFT framework’s key modules: red for the frozen previous-phase (Task $\tau_{t-1}$) components, green for the elements being trained in the current phase (Task $\tau_{t}$), and the purple box indicates the Acoustic Feature Transformation Network (AFT Network). The core AFT Network bridges old and new knowledge by aligning old acoustic features with the new model’s feature space, enabling the model to retain prior knowledge while adapting to new environmental sound categories. Additionally, the diagram’s plots show the feature space; the AFT Network updates old classes’ space after each task and incorporates new classes’ prototypes into it.}
  \label{fig:AFT}
  \vspace{-5mm}
\end{figure*}

To address this issue, we propose an Acoustic Feature Transformation (AFT) mechanism designed to mitigate acoustic feature variation and reduce catastrophic forgetting without relying on stored past data, making it ideal for privacy-sensitive scenarios. Inspired by the human brain’s memory mechanism, AFT builds connections between old and new tasks by aligning old feature representations into the new model's feature space. Its implementation involves three core steps: first, applying feature-level knowledge distillation to transfer key old-model information to the new one; second, transforming the old feature space into the new space for direct comparison to prevent feature variation; third, constructing a selectively compressed acoustic feature space via high-quality samples to maintain clear sound class boundaries. Our contributions are as follows: 
\begin{itemize}  [leftmargin=*]
\item We propose AFT, a novel mechanism that mitigates acoustic feature variation and reduces catastrophic forgetting without stored past data.
\item Inspired by human memory, AFT resolves dynamic learning’s feature inconsistency by cross-task alignment, strengthens sound boundaries with a selectively compressed acoustic feature space, and enhances model adaptability in real-world acoustic tasks.
\item Extensive experiments on two public datasets show AFT outperforms baseline models, achieving absolute accuracy gains of 3.7\% to 3.9\% and lower forgetting, validating its effectiveness.
\end{itemize}
\section{OUR METHOD}
\label{sec:format}

\subsection{Problem Formulation}
In this study, we aim to classify environmental sounds in stages and reduce forgetting. We assume the model learns a series of tasks, denoted as \{$\tau_1$, $\tau_2$, ..., $\tau_t$\}, in separate stages, where data from one task is not reused in others. For each task $\tau_t$, we are given a pair of inputs \{$x_t$, $y_t$\}, where $x_t$ is the environmental sound data and $y_t$ are the ground truth labels following the distribution $D_t$. Our goal is to minimize the cross-entropy loss~\cite{masana2022class} across all tasks after the entire sequence is learned.
\begin{align}
\mathcal{L}_{\text{ce}} = \sum_{(x_i,y_i) \in D_t} y_i\text{log}\frac{\text{exp}(o_i)}{\sum_{j \in D_t}\text{exp}(o_j)} 
\end{align}
where $o$ denotes the output logits of the model in the task $\tau_t$. 

\subsection{AFT Structure}
We present an overview of our proposed AFT in Figure.~\ref{fig:AFT}. In this section, we first introduce each module involved in the learning process, and then summarize the overall transformation framework employed in our method.

\noindent \textbf{Feature Distillation from Existing Model.}
To prevent catastrophic forgetting without storing old data, we keep the model weights from previous tasks to guide new learning. For task \(\tau_t\), we feed the input \(x_t\) into the model in previous stage and extract the feature \(f_{t-1}(x)\). We then compare \(f_{t-1}(x)\) with the feature \(f_t(x)\) from the current model. As a comparison, we calculate the \(\mathcal{L}_{\text{kfd}}\) as:
\begin{align}
\mathcal{L}_{\text{kfd}} = \parallel f_t(x) - f_{t-1}(x)  \parallel_2
\end{align}
where \(\parallel \cdot \parallel_2\) denotes the L2 norm. 
This process measures whether the new model still retains the knowledge of earlier tasks, allowing it to learn new tasks while maintaining past information.

\noindent \textbf{Acoustic Feature Transformation (AFT).}
Although Feature Distillation transfers knowledge from the old model to the new one, preserving all prior knowledge remains a challenge. This issue becomes more obvious when new sounds share high acoustic similarity with existing ones, as shown in Figure~\ref{fig:misalign}. As the new model adjusts its weights to learn new acoustic features, it fails to represent previously learned acoustic features with greater precision. This change distorts the old classes’ original feature space, further causes their representations to drift from the stable state, and weakens the model's ability to recognize old classes accurately. To fix this, we draw inspiration from the mechanism of the human brain's memory. The human brain consolidates old memories by establishing connections when learning new things. We apply an Acoustic Feature Transformation Network (AFT Network) $\mathcal{M}$, as illustrated in the purple box of Figure~\ref{fig:AFT}, to build a bridge between the features in the old and the new model. This bridge not only enables the new model to integrate novel class features but also leverages these new features to update prior knowledge of old classes. We update the transformation loss by:
\begin{align}
\mathcal{L}_{\text{trans}} = \parallel f_t(x) - \mathcal{M}(f_{t-1}(x))  \parallel_2
\end{align}
where \(\mathcal{M}(f_{t-1}(x)) \) denotes the output of the AFT Network when we feed the old feature \(f_{t-1}(x)\) into it. 

Furthermore, we store the acoustic feature distribution of each class after every task. However, when the model is updated, these stored distributions may no longer represent the old categories well in the new model. To solve this issue, we apply the transformation network to continuously update the old feature distributions, so they remain aligned with the current model.
Here, the original acoustic feature space captures the feature distribution of each trained category and is denoted as \(\mathcal{F}_{t-1}\). We first apply reparameterization to the data in \(\mathcal{F}_{t-1}\) to obtain reparameterized acoustic features. These reparameterized features are then fed into a transformation network. Subsequently, we calculate their logits directly through the feature classifier (FC). We define the loss function as:

\begingroup
\vspace{-5mm}
\begin{align}
\mathcal{L}_{\text{fs}} = \mathcal{L}_{\text{ce}}(g(\mathcal{M}(\mathcal{F}_{t-1})), y)
\end{align}
\vspace{-5mm}
\endgroup

where \(\mathcal{M}(\mathcal{F}_{t-1})\) is the transformed reparameterized acoustic features from the old task and $g$ denotes the fully connected network. 

\begin{figure}[t!]
  \centering
  \includegraphics[width=0.8\linewidth]{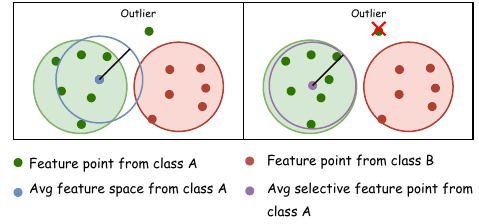}
  \vspace{-2mm}
  \caption{Selective Compression Acoustic Feature Space.}
  \vspace{-6mm}
\label{fig:selective_compression_acoustic_feature}
\end{figure}

\noindent \textbf{Selective Compression Acoustic Feature Space.} As mentioned, retaining old features is crucial for preserving past knowledge when no historical data is stored. A common method~\cite{gomez2025exemplar} is to compute the average of all training data features for each category and use this average as the category's feature representation. However, this method depends heavily on the quality of the training data. If the data for a category is scattered or contains many errors, the average may not capture the true characteristics of that category and may even overlap with other categories, as shown in the left part of Figure~\ref{fig:selective_compression_acoustic_feature}. To address this, we propose a solution that carefully selects high-quality training samples—{\it focuses on data points that best represent each category}. As illustrated on the right part of Figure~\ref{fig:selective_compression_acoustic_feature},  we identify and exclude outliers by comparing model outputs to ground truth. In addition, we maintain the original radiuses in the feature space rather than those of selected features, enhancing the robustness of the acoustic feature space.

\noindent \textbf{Training Overview.}  
Our AFT approach addresses the core challenges of feature misalignment for acoustically sound classes. It starts by using knowledge distillation to transfer discriminative information from a pre-trained model to the new model using the current data. It then introduces the AFT Network to build a relational connection between past and present knowledge, directly alleviating feature drift and avoiding confusion between features of similar classes such as \enquote{jackhammer} and \enquote{drilling}. Finally, an adaptive filter refines the preserved feature distribution by pruning abnormal prototypes, sharpening class boundaries and enhancing model stability. The overall loss function is defined as:
\begingroup
\vspace{-2mm}
\begin{align}
Loss = \mathcal{L}_{\text{ce}} + \alpha\mathcal{L}_{\text{kfd}} + \beta\mathcal{L}_{\text{trans}} + \gamma\mathcal{L}_{\text{fs}}
\end{align}
\vspace{-6mm}
\endgroup

\noindent where $\alpha$, $\beta$, $\gamma$ are the hyperparameters that balance the different loss components. Here, we run grid search on $\alpha$ = \{0.1,1,1.5,2\}, $\beta$=\{1,5,15,18,20\} and $\gamma$=\{1,5,15,18,20\} respectively.

\section{Experimental Settings}
\label{sec:pagestyle}

\subsection{Datasets.}
In this study, we evaluate our method using two datasets to show its effectiveness in environmental sound classification. Both datasets contain environmental sounds, but they differ in acoustic characteristics and sound length. \textbf{UrbanSound8K}~\cite{salamon2014dataset} has 8,732 carefully labeled sound snippets. Each snippet lasts up to 4 seconds and covers 10 urban sound classes. \textbf{DCASE 2019 Task 1}~\cite{Mesaros2019} is designed for acoustic scene classification. Its development set consists of 10-second audio clips from 10 scenes.

\subsection{Backbone Network.}
We first extract Mel-frequency cepstrum coefficients (MFCCs) from the audio signals and input them into the model. We adopt the TC-ResNet-8~\cite{choi19_interspeech} as backbone, whose \((1\times t)\) temporal convolution and frequency-untied design inherently preserve formant locality and streaming capability, aligning with exemplar-free CIL constraints on edge devices. The network consists of three residual blocks, each containing 1D temporal convolutional layers, batch normalization, and ReLU activation. The channels are set to 16, 24, 32, and 48, including the initial convolutional layer.

\subsection{Implementation Details.}
We convert the original audio to a 16 kHz sampling rate in mono. For the UrbanSound8K dataset, we use only the first 3 seconds of each clip, while for the DCASE 2019 Task 1 dataset, we use the full 10-second segments. We extract 40-dimensional Mel-frequency cepstral coefficients from audio using the log-mel spectrogram. The classification network is optimized with the Adam algorithm \cite{kingma2014adam} at a learning rate of $1 \times 10^{-3}$. The batch size is set to 128, and the model is trained for 50 epochs. We pre-train the model on 5 categories and then treat the remaining categories as new tasks, resulting in a total of 5 tasks.
\begin{table}[!t]
  \caption{Comparison between AFT with recent state-of-the-art training strategies on two datasets, using the TCResNet-8 model as a common testbed to ensure fair evaluation.}
  \label{tab:main_result}
  \centering
\scalebox{0.95}{
    \begin{tabular}{lcccc}  
        \toprule
        
        \multirow{2}{*}{\textbf{Method}}  & \multicolumn{2}{c}{\textbf{UrbanSound8K}} & \multicolumn{2}{c}{\textbf{DCASE 2019 Task 1}}\\
        \cmidrule{2-3} \cmidrule{4-5}  
        & \textbf{ACC(\%)}           & \textbf{BWT}    & \textbf{ACC(\%)}           & \textbf{BWT}  \\
        \midrule
        Finetune              &26.700    &-0.368      &22.900       & -0.267                     \\
        \midrule
        EWC~\cite{kirkpatrick2017overcoming}        &29.284           & -0.358       &23.472     & -0.264  \\
        SI~\cite{zenke2017continual}             &42.267         & -0.264    &26.802         & -0.233 \\
        LWF~\cite{li2017learning}        &52.285       &-0.198 &46.965 &-0.097             \\
        LDC~\cite{gomez2025exemplar}         & 56.703        & -0.157        &48.867 & -0.104 \\
        \textbf{AFT}            &\textbf{60.464}   &\textbf{-0.147}   &\textbf{52.762}     &\textbf{-0.077}                   \\
        \midrule
        Joint               &93.204     &- &66.725    &-               \\
        \bottomrule
    \end{tabular}
}
\vspace{-6mm}
\end{table}
\subsection{Baselines.}
We set two simple methods as our performance bounds. Fine-tuning serves as the lower bound, while joint training is the upper bound. In addition, we compare our method with four classic and powerful exemplar-free approaches.
\begin{itemize} [leftmargin=*]


\item \textbf{Synaptic Intelligence (SI)}~\cite{zenke2017continual} assigns a heavy score to each synapse to measure its role in past tasks. During new task training, it penalizes changes to key weights to protect old knowledge.

\item \textbf{Elastic Weight Consolidation (EWC)}~\cite{kirkpatrick2017overcoming} adds a quadratic penalty to the loss function based on the fisher information matrix, restricting updates to weights critical for prior tasks.

\item \textbf{Learning without Forgetting (LWF)}~\cite{li2017learning} uses a knowledge distillation loss to retain information from old tasks while learning new ones, all without using old task data. 

\item \textbf{Learnable Drift Compensation (LDC)}~\cite{gomez2025exemplar} accumulates category prototypes to counter semantic drift in the feature space, thus addressing catastrophic forgetting.
\end{itemize}

\subsection{Metrics.} 
We evaluate our approach using two metrics. We utilize Average Accuracy (ACC) and Backward Transfer (BWT)~\cite{lopez2017gradient}. The ACC reflects the overall performance across all tasks after training, while the BWT assesses how learning new tasks affects the accuracy of previous tasks with a negative value indicating the model has forgotten a portion of its prior knowledge. 

\section{Results and Analysis}
\label{sec:typestyle}

\subsection{Main Results}
As shown in Table \ref{tab:main_result}, for the CIL task on two datasets, different methods yield varying results, resulting from their distinct designs for retaining old knowledge. Finetuning, which serves as the lower bound for baseline comparison, shows severe forgetting on both datasets because it lacks any mechanism to retain the features of old classes.
	
	The table shows two exemplar-free methods with relatively better performance have made improvements. LWF adopts a dual-loss constraint, combining cross-entropy loss for new tasks and knowledge distillation loss from old model outputs, achieving 52.29\% accuracy (BWT of -0.198) on UrbanSound8K and 46.97\% accuracy (BWT of -0.097) on DCASE 2019 Task 1. However, it only retains old knowledge indirectly through old model outputs and does not adjust the feature space, leading to old feature misalignment and lost recognition cues when handling acoustically similar classes (e.g., drilling and jackhammer) in both datasets. LDC performs better on both datasets by training a feature extractor via self-supervision and learning feature space drift compensation, achieving 56.70\% accuracy (BWT of -0.157) on UrbanSound8K and 48.87\% accuracy (BWT of -0.104) on DCASE 2019 Task 1. Since this method relies on a good feature extractor and both datasets have relatively small sizes, its drift compensation fails to retain old knowledge effectively.
	
	Our proposed AFT method demonstrates better performance in the results. On UrbanSound8K, where the dataset size is small and some class features are similar, AFT enhances the boundaries between similar classes by mapping the feature space and selecting high-quality samples, ultimately achieving an accuracy of 60.46\% with a BWT of -0.147. On DCASE 2019 Task 1, it further adapts to more complex acoustic environments, ultimately achieving an accuracy of 52.76\% with a BWT of -0.077. 
\begin{figure}[t!]
  \centering
  \includegraphics[width=1.0\linewidth]{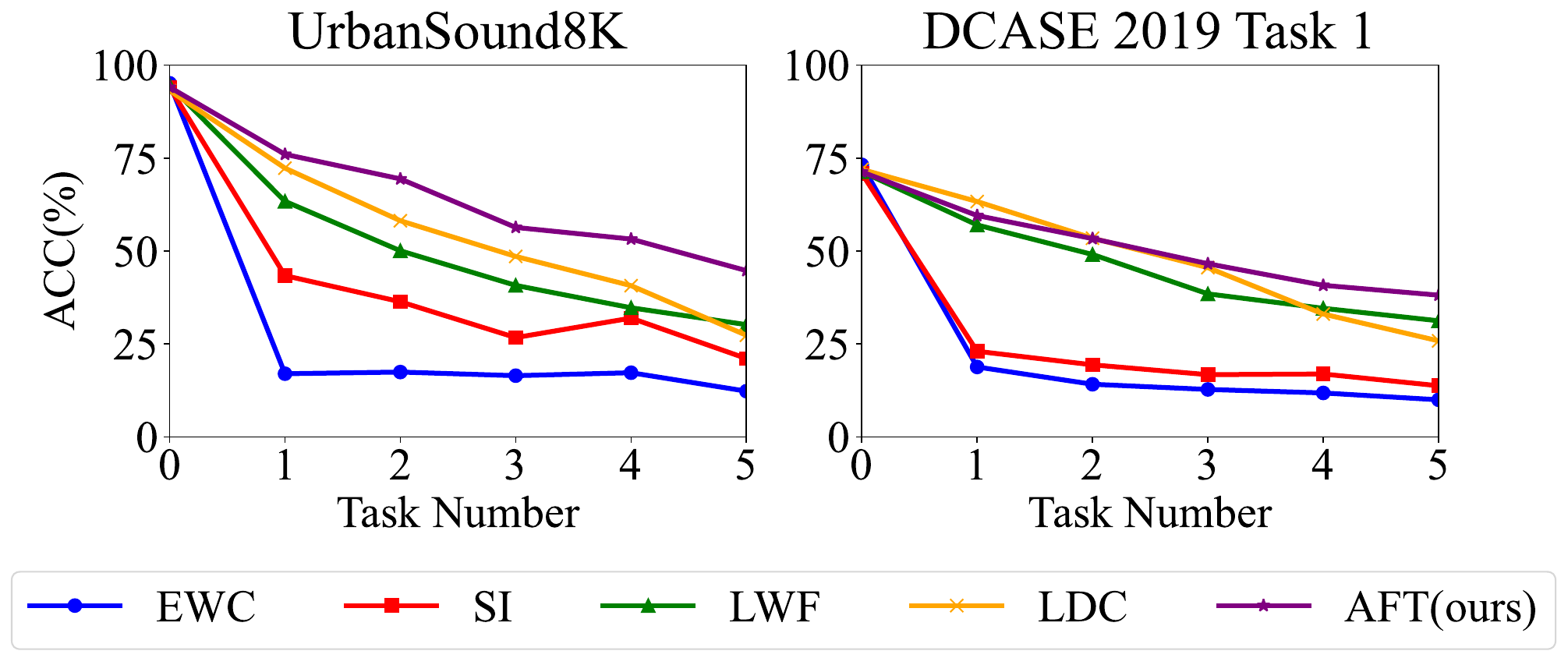}
  \vspace{-4mm}
  \caption{The ACC (\%) comparison of the proposed AFT approach and other competitive baselines on two datasets.}
  \label{fig:accuray_curve}
  \vspace{-3mm}
\end{figure}

To clarify our results, Figure~\ref{fig:accuray_curve} displays the classification accuracy of our method on two datasets. Notably, all methods had similar initial accuracy. As the number of tasks increased, accuracy dropped for all methods. However, AFT achieved the best results in later stages, with a relatively flat forgetting curve. This shows that our method balances knowledge retention and acquisition better in ESC.

\subsection{Ablation Study}
\textbf{Impact of Acoustic Feature Transformation (AFT).}
Table~\ref{tab:ablation} shows that adding AFT to the baseline (Base + AFT) boosts UrbanSound8K accuracy from 57.47\% to 60.17\% (BWT from -0.159 to -0.148) and DCASE 2019 Task 1 accuracy from 50.69\% to 52.40\%, confirms that aligning old and new features through transformation helps preserve earlier knowledge while learning new tasks.

\noindent \textbf{Impact of Selective Compression Acoustic Feature Space (POS).}
Replacing the general feature space transformation with our selective compression method (Base + AFT + POS) brings an additional boost to 60.46\% on UrbanSound8K and 52.76\% on DCASE 2019 Task 1, and further reduces forgetting (BWT from -0.148 to -0.147 on UrbanSound8K and -0.077 to -0.0766 on DCASE 2019 Task 1). Although the numerical increase may appear slight, it reflects the method’s ability to filter out low-quality features, ensuring clearer class boundaries and more stable learning across tasks.
\begin{table}[t!]
    \vspace{-1mm}  
  \caption{Ablation study of different components}
  \vspace{-1mm}
  \label{tab:ablation}
  \centering
  \scalebox{0.85}{
    \begin{tabular}{lcccc} 
      \toprule
      \multirow{2}{*}{\textbf{Method}} & \multicolumn{2}{c}{\textbf{UrbanSound8K}} & \multicolumn{2}{c}{\textbf{DCASE 2019 Task 1}} \\
      \cmidrule(lr){2-3} \cmidrule(lr){4-5}
      & \textbf{ACC(\%)} & \textbf{BWT} & \textbf{ACC(\%)} & \textbf{BWT} \\
      \midrule
      Base & 57.474 & -0.15853 & 50.688 & -0.08508 \\
      Base+AFT & 60.167 & -0.14783 & 52.396 & -0.07698 \\
      \rowcolor{gray!20} Base+AFT+POS & 60.464 & -0.14726 & 52.762 & -0.07658 \\
      \bottomrule
    \end{tabular}
  }
  \vspace{-2mm}
\end{table}
\subsection{Feature Space Visualization with t-SNE}
\begin{figure}[h]
  \vspace{-1mm}
  \centering
  \begin{subfigure}[t]{0.32\linewidth}
    \includegraphics[width=\linewidth]{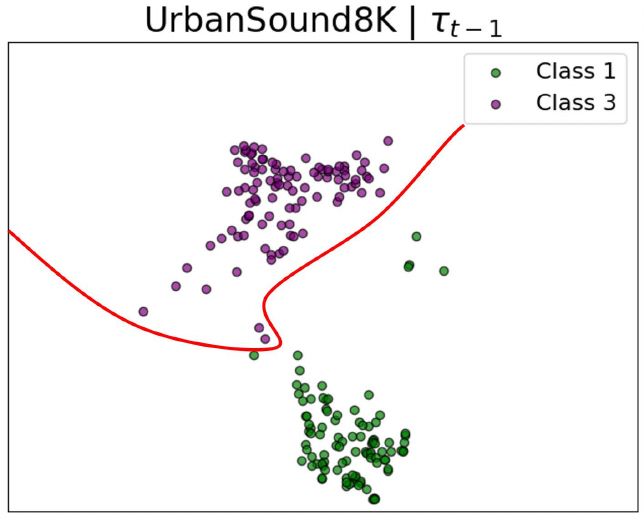}
    \caption{}\label{fig:1}
  \end{subfigure}
  \begin{subfigure}[t]{0.32\linewidth}
    \includegraphics[width=\linewidth]{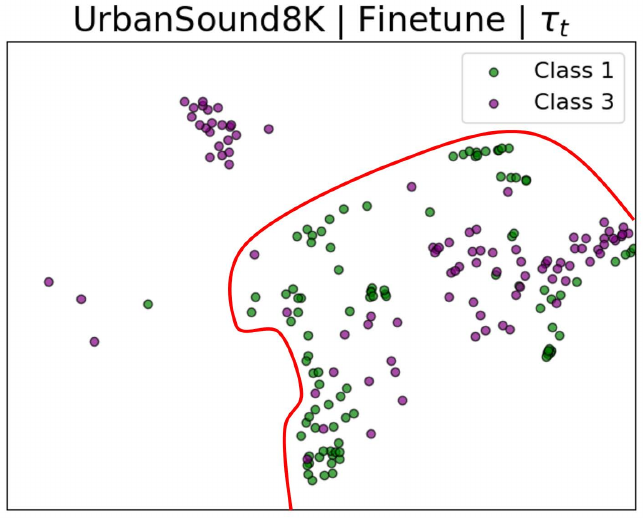}
    \caption{}\label{fig:finetune_2}
  \end{subfigure}
  \begin{subfigure}[t]{0.32\linewidth}
    \includegraphics[width=\linewidth]{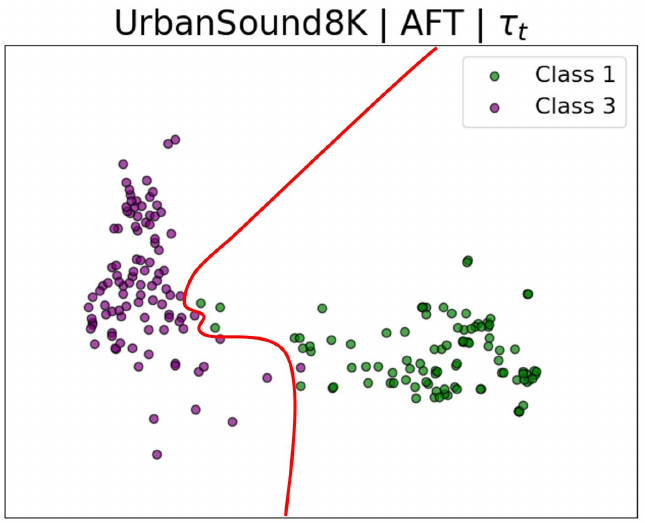}
    \caption{}\label{fig:AFT_2}
  \end{subfigure}
\vspace{-2mm}
  \caption{The t-SNE visualization of two classes from UrbanSound8K across different stages: (a) the original feature space, (b) the feature space after fine-tuning, and (c) the feature space after applying AFT.}
  \label{fig:size}
  \vspace{-6mm}
\end{figure}

	Figure~\ref{fig:size} illustrates the change of feature boundaries during incremental learning, showing the original distribution, the distortion after fine-tuning, and the adjustment with AFT for two classes. Compared with (a), (b) shows that fine-tuning shifts the features and reduces boundary clarity. In (c), AFT realigns the old features into the new space while maintaining separation. These results suggest that AFT helps retain class distinctions across tasks.
\section{Conclusion}
\label{sec:majhead}

This study presents Acoustic Feature Transformation (AFT), an innovative exemplar-free continual learning method designed to balance between retaining existing knowledge and adapting to new tasks in the domain of acoustic feature adaptation. AFT achieves this by projecting old features into a new model space to create a memory link between old and new tasks, while also selectively filtering the feature space. This approach allows AFT to mitigate the issue of catastrophic forgetting without history data, making it suitable for scenarios where data privacy is a primary concern. 
Our evaluation on two public datasets shows that AFT outperforms existing CIL methods. 
These findings underscore the potential of AFT to enhance the performance of acoustic feature adaptation in practical scenarios while maintaining high efficiency.




\vfill\pagebreak

\footnotesize
\bibliographystyle{IEEEbib}
\bibliography{strings}

\begin{thebibliography}{10}

\bibitem{piczak2015esc}
Karol~J Piczak,
\newblock ``{ESC: }dataset for environmental sound classification,''
\newblock in {\em Proc. ACM International Conference on Multimedia}, 2015, pp. 1015--1018.

\bibitem{10.1016/j.asoc.2024.112619}
Jia-Wei Chang, Hao-Shang Ma, and Zhong-Yun Hu,
\newblock ``Multi-level transfer learning using incremental granularities for environmental sound classification and detection,''
\newblock {\em Appl. Soft Comput.}, vol. 169, no. C, Jan. 2025.

\bibitem{yin2025exploring}
Han Yin, Jisheng Bai, Yang Xiao, Hui Wang, Siqi Zheng, Yafeng Chen, Rohan~Kumar Das, Chong Deng, and Jianfeng Chen,
\newblock ``Exploring text-queried sound event detection with audio source separation,''
\newblock in {\em Proc. IEEE International Conference on Acoustics, Speech and Signal Processing (ICASSP)}, 2025, pp. 1--5.

\bibitem{xiao2024mixstyle}
Yang Xiao, Han Yin, Jisheng Bai, and Rohan~Kumar Das,
\newblock ``Mixstyle based domain generalization for sound event detection with heterogeneous training data,''
\newblock {\em arXiv preprint arXiv:2407.03654}, 2024.

\bibitem{xiao2024wilddesed}
Yang Xiao and Rohan~Kumar Das,
\newblock ``Wilddesed: an llm-powered dataset for wild domestic environment sound event detection system,''
\newblock in {\em Proc. Detection and Classification of Acoustic Scenes and Events Workshop (DCASE)}, 2024.

\bibitem{masana2022class}
Marc Masana, Xialei Liu, Bart{\l}omiej Twardowski, Mikel Menta, Andrew~D Bagdanov, and Joost Van De~Weijer,
\newblock ``Class-incremental learning: survey and performance evaluation on image classification,''
\newblock {\em IEEE Transactions on Pattern Analysis and Machine Intelligence}, vol. 45, no. 5, pp. 5513--5533, 2022.

\bibitem{wang2024comprehensive}
Liyuan Wang, Xingxing Zhang, Hang Su, and Jun Zhu,
\newblock ``A comprehensive survey of continual learning: theory, method and application,''
\newblock {\em IEEE Transactions on Pattern Analysis and Machine Intelligence}, 2024.

\bibitem{zhuang2022acil}
Huiping Zhuang, Zhenyu Weng, Hongxin Wei, Renchunzi Xie, Kar-Ann Toh, and Zhiping Lin,
\newblock ``{ACIL:} analytic class-incremental learning with absolute memorization and privacy protection,''
\newblock {\em Advances in Neural Information Processing Systems}, vol. 35, pp. 11602--11614, 2022.

\bibitem{pandey2024class}
Yang Xiao, Peng Tianyi, Rohan~Kumar Das, Yuchen Hu, and Huiping Zhuang,
\newblock ``{A}nalytic{KWS}: Towards exemplar-free analytic class incremental learning for small-footprint keyword spotting,''
\newblock in {\em Findings of the Association for Computational Linguistics}, 2025, pp. 14147--14158.

\bibitem{xiao25c_interspeech}
Yang Xiao and Rohan~Kumar Das,
\newblock ``{Listen, Analyze, and Adapt to Learn New Attacks: An Exemplar-Free Class Incremental Learning Method for Audio Deepfake Source Tracing},''
\newblock in {\em {Interspeech}}, 2025, pp. 1563--1567.

\bibitem{zhuang2024ds}
Huiping Zhuang, Run He, Kai Tong, Ziqian Zeng, Cen Chen, and Zhiping Lin,
\newblock ``Ds-al: A dual-stream analytic learning for exemplar-free class-incremental learning,''
\newblock in {\em Proc. the AAAI Conference on Artificial Intelligence}, 2024, vol.~38, pp. 17237--17244.

\bibitem{xiao2022continual}
Yang Xiao, Xubo Liu, James King, Arshdeep Singh, Eng~Siong Chng, Mark~D. Plumbley, and Wenwu Wang,
\newblock ``Continual learning for on-device environmental sound classification,''
\newblock in {\em Proc. Detection and Classification of Acoustic Scenes and Events Workshop (DCASE)}, 2022.

\bibitem{wang2019continual}
Zhepei Wang, Cem Subakan, Efthymios Tzinis, Paris Smaragdis, and Laurent Charlin,
\newblock ``Continual learning of new sound classes using generative replay,''
\newblock in {\em Proc. IEEE Workshop on Applications of Signal Processing to Audio and Acoustics (WASPAA)}, 2019, pp. 308--312.

\bibitem{xiao2024ucil}
Yang Xiao and Rohan~Kumar Das,
\newblock ``{UCIL}: An unsupervised class incremental learning approach for sound event detection,''
\newblock in {\em Proc. IEEE International Conference on Acoustics, Speech and Signal Processing (ICASSP)}, 2025.

\bibitem{yang2023dual}
Zhao Yang, Dianwen Ng, Xizhe Li, Chong Zhang, Rui Jiang, Wei Xi, Yukun Ma, Chongjia Ni, Jizhong Zhao, Bin Ma, et~al.,
\newblock ``Dual-memory multi-modal learning for continual spoken keyword spotting with confidence selection and diversity enhancement,''
\newblock in {\em Proc. Interspeech}, 2023.

\bibitem{rk}
Yang Xiao, Nana Hou, and Eng~Siong Chng,
\newblock ``{Rainbow Keywords: Efficient Incremental Learning for Online Spoken Keyword Spotting},''
\newblock in {\em Proc. Interspeech}, 2022, pp. 3764--3768.

\bibitem{yang2024improving}
Muqiao Yang, Umberto Cappellazzo, Xiang Li, and Bhiksha Raj,
\newblock ``Improving continual learning of acoustic scene classification via mutual information optimization,''
\newblock in {\em Proc. IEEE International Conference on Acoustics, Speech and Signal Processing (ICASSP)}. IEEE, 2024, pp. 7105--7109.

\bibitem{peng2024dark}
Tianyi Peng and Yang Xiao,
\newblock ``Dark experience for incremental keyword spotting,''
\newblock in {\em Proc. IEEE International Conference on Acoustics, Speech and Signal Processing (ICASSP)}, 2025.

\bibitem{zhang2024remember}
Xiaohui Zhang, Jiangyan Yi, Chenglong Wang, Chu~Yuan Zhang, Siding Zeng, and Jianhua Tao,
\newblock ``What to remember: Self-adaptive continual learning for audio deepfake detection,''
\newblock in {\em Proc. the AAAI Conference on Artificial Intelligence}, 2024, vol.~38, pp. 19569--19577.

\bibitem{berg2021continual}
Jan Berg and Konstantinos Drossos,
\newblock ``Continual learning for automated audio captioning using the learning without forgetting approach,''
\newblock in {\em Proc. Detection and Classification of Acoustic Scenes and Events Workshop (DCASE)}, 2021, pp. 140--144.

\bibitem{yue2024mmal}
Xianghu Yue, Xueyi Zhang, Yiming Chen, Chengwei Zhang, Mingrui Lao, Huiping Zhuang, Xinyuan Qian, and Haizhou Li,
\newblock ``{MMAL:} multi-modal analytic learning for exemplar-free audio-visual class incremental tasks,''
\newblock in {\em Proc. ACM International Conference on Multimedia}, 2024, pp. 2428--2437.

\bibitem{xiao2024configurable}
Yang Xiao and Rohan~Kumar Das,
\newblock ``Where's that voice coming? {Continual} learning for sound source localization,''
\newblock in {\em Proc. IEEE International Conference on Multimedia and Expo (ICME)}, 2024.

\bibitem{10447952}
Manjunath Mulimani and Annamaria Mesaros,
\newblock ``Class-incremental learning for multi-label audio classification,''
\newblock in {\em Proc. IEEE International Conference on Acoustics, Speech and Signal Processing (ICASSP)}, 2024, pp. 916--920.

\bibitem{gomez2025exemplar}
Alex Gomez-Villa, Dipam Goswami, Kai Wang, Andrew~D Bagdanov, Bartlomiej Twardowski, and Joost van~de Weijer,
\newblock ``Exemplar-free continual representation learning via learnable drift compensation,''
\newblock in {\em European Conference on Computer Vision}. Springer, 2025, pp. 473--490.

\bibitem{salamon2014dataset}
Justin Salamon, Christopher Jacoby, and Juan~Pablo Bello,
\newblock ``A dataset and taxonomy for urban sound research,''
\newblock in {\em Proc. ACM International Conference on Multimedia}, 2014.

\bibitem{Mesaros2019}
Annamaria Mesaros, Toni Heittola, and Tuomas Virtanen,
\newblock ``Acoustic scene classification in dcase 2019 challenge: Closed and open set classification and data mismatch setups,''
\newblock in {\em Proc. Detection and Classification of Acoustic Scenes and Events Workshop (DCASE)}, 2019, pp. 164--168.

\bibitem{choi19_interspeech}
Seungwoo Choi, Seokjun Seo, Beomjun Shin, Hyeongmin Byun, Martin Kersner, Beomsu Kim, Dongyoung Kim, and Sungjoo Ha,
\newblock ``Temporal convolution for real-time keyword spotting on mobile devices,''
\newblock in {\em Interspeech}, 2019.

\bibitem{kingma2014adam}
Diederik~P. Kingma and Jimmy Ba,
\newblock ``Adam: {A} method for stochastic optimization,''
\newblock in {\em Proc. International Conference on Learning Representations {(ICLR)}}, 2015.

\bibitem{kirkpatrick2017overcoming}
James Kirkpatrick, Razvan Pascanu, Neil Rabinowitz, Joel Veness, Guillaume Desjardins, Andrei~A Rusu, Kieran Milan, John Quan, Tiago Ramalho, Agnieszka Grabska-Barwinska, et~al.,
\newblock ``Overcoming catastrophic forgetting in neural networks,''
\newblock {\em Proc. the national academy of sciences}, vol. 114, no. 13, pp. 3521--3526, 2017.

\bibitem{zenke2017continual}
Friedemann Zenke, Ben Poole, and Surya Ganguli,
\newblock ``Continual learning through synaptic intelligence,''
\newblock in {\em Proc. International Conference on Machine Learning}, 2017, pp. 3987--3995.

\bibitem{li2017learning}
Zhizhong Li and Derek Hoiem,
\newblock ``Learning without forgetting,''
\newblock {\em IEEE Transactions on Pattern Analysis and Machine Intelligence}, vol. 40, no. 12, pp. 2935--2947, 2017.

\bibitem{lopez2017gradient}
David Lopez-Paz and Marc'Aurelio Ranzato,
\newblock ``Gradient episodic memory for continual learning,''
\newblock {\em Advances in neural information processing systems}, vol. 30, 2017.

\end{thebibliography}

\end{document}